\documentstyle[emulateapj,apjfonts,graphics]{article}

\slugcomment{Accepted for publication in the Astrophysical Journal}
\lefthead{WECHSLER, GROSS, PRIMACK, BLUMENTHAL, \& DEKEL }
\righthead{REDSHIFT DISTRIBUTION OF $z\sim 3$ GALAXIES}

\def\capt{\footnotesize}

\def\eg{{\it e.g.}}
\def\etal{{\it et al.}}
\def\hmpc{\mbox{$h^{-1}$ Mpc}}
\def\hmsun{\mbox{$h^{-1}$ $M_\odot$}}
\def\hkpc{\mbox{$h^{-1}$ kpc}}
\def\kmsmpc{\mbox{km s$^{-1}$ Mpc$^{-1}$}}
\def\Mh{M_{\mathrm{h}}}
\def\sigmah{\sigma_{\rm h}}
\def\sigmap{\sigma_{\rm p}}
\def\sigmahp{\sigma^{\rm h}_{\rm p}}

\def\msun{\mbox{$M_\odot$}}
\def\hmsun{\mbox{$h^{-1}$ $M_\odot$}}
 
\def\delg{\delta_{\rm g}}
\def\delc{\delta_{\rm c}}
\def\delm{\delta_{\rm m}}
\def\lcdm{$\Lambda$CDM}

\def\contcaption#1{%
  \addtocounter{figure}{-1}
  \renewcommand{\thefigure}{\arabic{figure}. (continued)}
  \caption{#1}
  \renewcommand{\thefigure}{\arabic{figure}.}
}

\begin{document}

\title{IMPLICATIONS OF SPIKES IN THE REDSHIFT DISTRIBUTION OF $z\sim3$ GALAXIES}

\author{Risa H. Wechsler\altaffilmark{1}, 
        Michael A. K. Gross\altaffilmark{1,2}, \\
        Joel R. Primack\altaffilmark{1},
        George R. Blumenthal\altaffilmark{3,4},
     \& Avishai Dekel\altaffilmark{1,3,4}
       }

\altaffiltext{1}{Physics Department, University of California, Santa Cruz, 
                 CA 95064}
\altaffiltext{2}{NASA/Goddard Space Flight Center, Code 931, 
		 Greenbelt, MD 20771}

\altaffiltext{3}{Astronomy Department, University of California, Santa Cruz, 
                 CA 95064}
\altaffiltext{4}{Racah Institute of Physics, The  Hebrew University,
                 Jerusalem 91904, Israel}

\begin{abstract}

We address the high peaks found by Steidel \etal\ (1998) in the redshift
distribution of ``Lyman-break'' objects (LBOs) at redshift $z\simeq 3$.
The highest spike represents a relative overdensity of 2.6 in the 
distribution of LBOs in pixels of comoving size $\sim 10$ \hmpc. 
We examine the likelihood of such a spike in the redshift distribution 
within a suite of models for the evolution of structure in the Universe, 
including models with $\Omega=1$ (SCDM and CHDM) and with 
$\Omega_0 = 0.4-0.5$ (\lcdm\ and OCDM). 
Using high-resolution dissipationless 
N-body simulations, we analyze deep pencil-beam surveys 
from these models in the same way that they are actually observed,
identifying LBOs with the most massive dark matter halos. 
We find that all the models (with SCDM as a marginal exception)
have a substantial probability of producing spikes similar to those observed,
because the massive halos are much more clumped than the 
underlying matter -- i.e., they are biased. 
Therefore, the likelihood of such a spike is not a good discriminator 
among these models.  
We find in these models that the mean biasing parameter $b$
of LBOs with respect to dark matter varies within a range 
$b\simeq 2-5$ on a scale of $\sim 10$ \hmpc. 
However, all models show considerable dispersion in their biasing, with the 
local biasing parameter reaching values as high as ten.
We also compute the two-body correlation functions of 
LBOs predicted in these models.
The LBO correlation functions are less steep than galaxies today
($\gamma\approx-1.4$), but show similar or slightly longer 
correlation lengths.

\end{abstract}

\subjectheadings{
cosmology: theory --- 
cosmology: observation ---
dark matter --- 
galaxies: formation --- 
galaxies: clustering ---
large-scale structure of universe}

\section{INTRODUCTION}
\label{sec:intro}

Deep pencil redshift surveys of galaxies can reveal significant
information about the clustering of galaxies and the presence of a
scale in the structure of the Universe (e.g., Broadhurst \etal\ 1992, Dekel
\etal\ 1991).  Recently, Steidel \etal\ (1998; hereafter S98) discovered
that the redshift distribution of ``Lyman-break" objects (LBOs) in a
pencil beam reveals a large ``spike" in the LBO
distribution near $z\simeq 3.1$. This spike corresponds to a
fractional overdensity of LBOs of a few hundred percent over a
comoving scale of order $\sim 10$ \hmpc.

At a first glance, the occurrence of such a dramatic spike seems
surprising. This peak suggests substantial nonlinear clustering on
rather large scales --- scales that are normally considered linear at
such early epochs.  Indeed, S98 argue that 
low-$\Omega$ 
models of the Universe require a biasing parameter $b\simeq 2-4$ to
produce such spikes, while models with $\Omega=1$ require 
even higher biasing. In doing their analysis, S98 attempted a
comparison with theoretical scenarios in the ``theoretical plane",
corresponding to the matter distribution in 
real space (as opposed to redshift space) and to epochs of
linear evolution. In particular, they translated the data step by step
all the way to the linear-fluctuation power spectra, by performing an
analytic calculation based on the latest wisdom regarding galaxy
``biasing", redshift distortions, and nonlinear gravitational
evolution.  The approximations involved in this analysis are
somewhat crude, yielding uncertain conclusions.

Subsequently, Bagla (1997) compared the observed spike to the results
of pencils in cold dark matter (CDM) numerical simulations. By
defining galaxies as regions of high matter density, 
he found that high spikes were not rare events in such models.
The simulations we use for our analysis, unlike those used in Bagla (1997),  
are of sufficiently high resolution to resolve individual 
galaxy halos at very high redshift.  
In the present paper, we identify $z\sim3$ 
LBOs with the most massive halos in our simulations at that redshift. 

In this paper, we compare the spike in the LBOs of the S98 data 
to what is expected in four different cosmological models. We do this
comparison in the ``observational plane", by observing our numerical
simulations just as an observer would. By comparing 
a suite of different models, we can determine whether the likelihood of a
large spike differs greatly among different cosmological models. 
In relation to the analysis of S98, we examine the biasing of LBO halos 
compared to the underlying matter. 
We finally compute the two-point correlation functions of halos in these
models. 
In a Note added, we discuss the relation of our work to other recent 
papers, and in the Appendix we give a simple analytic treatment of bias that 
helps explain the results from our simulations.

\section{OBSERVATIONAL DATA}
\label{sec:data}

\begin{figure*}[htb!]
\begin{center}
\resizebox{0.47\textwidth}{!}{\rotatebox{-90}{\includegraphics{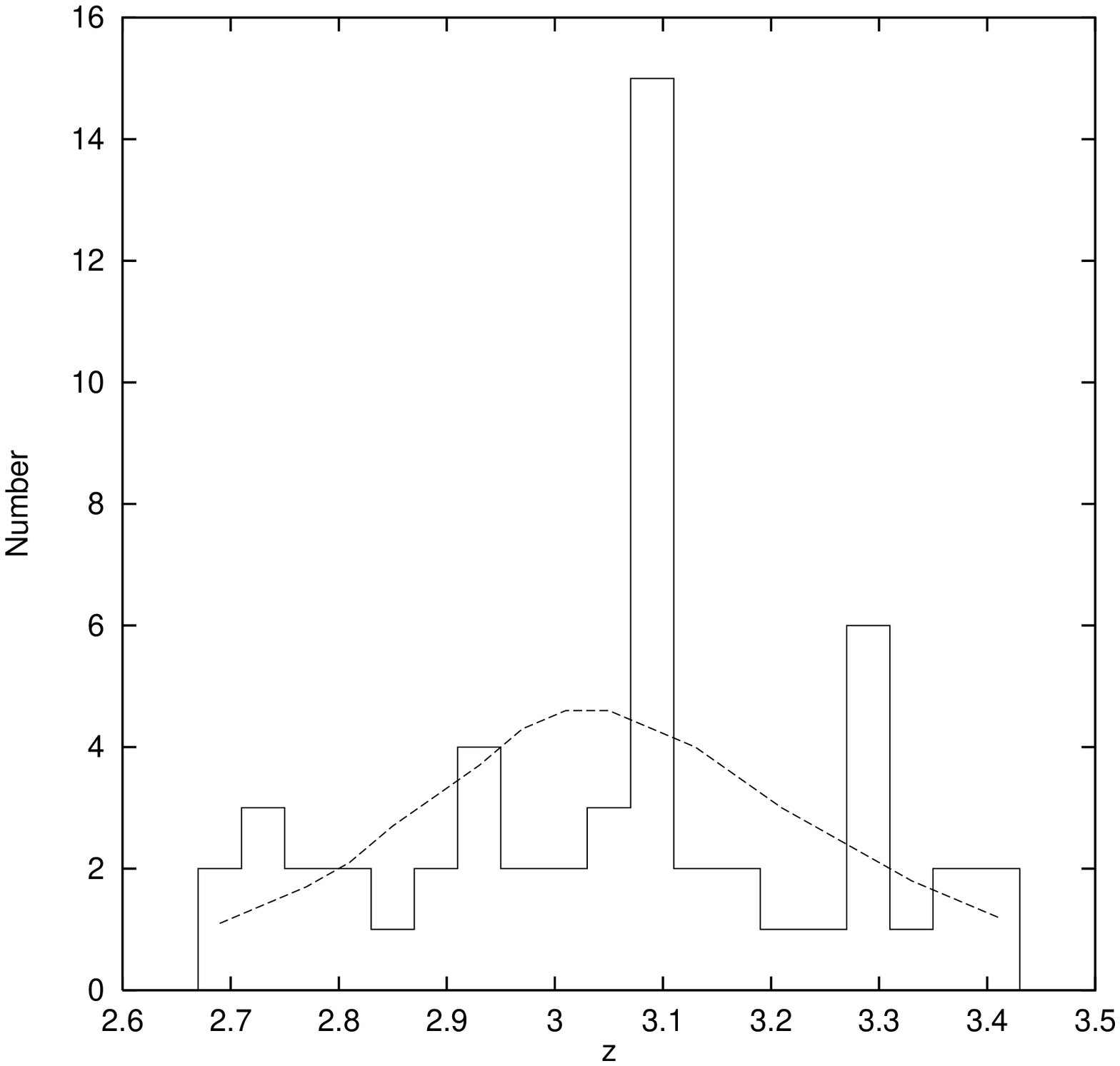}}}\hfill
\resizebox{0.48\textwidth}{!}{\rotatebox{-90}{\includegraphics{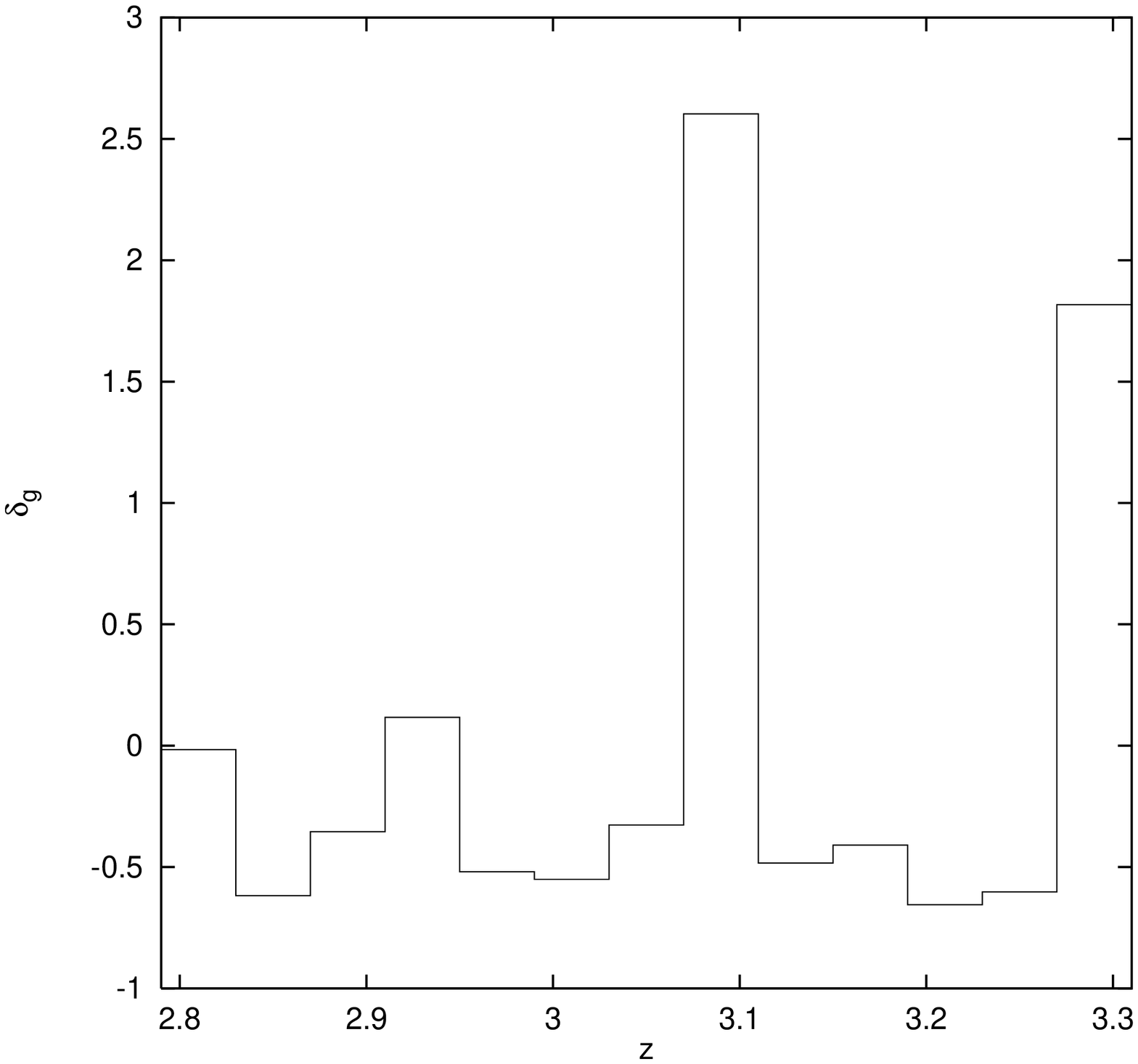}}}
\end{center}
\caption{\capt
{\em Left panel}: Redshift histogram from Steidel \etal\ (1997) showing 
  the number
  of LBOs in each redshift bin. Also shown is their estimated
  selection function as a function of redshift. 
{\em Right panel}: Redshift histogram
  of LBOs divided by the selection function to better reflect the true
  underlying distribution in redshift space of LBOs. We show only the
  the 13 central pixels, and instead of the corrected number of LBOs,
  we plot the relative overdensity of LBOs compared to their average
  number in a pixel.}
  \label{fig:data}
\end{figure*}

Figure~\ref{fig:data}(a) shows the raw number counts by S98 as a function 
of redshift, and their estimated selection function.
The redshift bins, of width $\Delta z=0.04$,
represent three-dimensional pixels that are almost rectangular. Their
angular dimensions are $\Delta\theta=8'.7$ and $\Delta\phi=17'.6$. 
We limit our analysis of the data to the redshift range $2.79<z<3.31$ 
within which the selection function is higher than 2. This is about 
43\% of its value at its maximum near $z=3.02$. This 
cut leaves us with 13 pixels.

{\footnotesize
\begin{deluxetable}{cccccccccc}
\tablenum{1}
\tablecolumns{10}
\tablecaption{Summary of Parameters for Four Cosmological Models}
\tablehead{
\colhead{Model} 	 & \colhead{$z$ range} &
\colhead{$N_{\rm pixels}$} & 
\colhead{Pixel Volume} & 
\colhead{Mass Threshold} &
\colhead{$p$}            & \colhead{$q$} &
\colhead{$P_{1}$}	 & \colhead{$P_{2}$}	 & 
\colhead{Mean $b$} \\
\colhead{}	& \colhead{}& 
\colhead{}& 
\colhead{($h^{-1}$ Mpc)$^3$} & 
\colhead{($h^{-1}$ \msun)} &
\colhead{}& \colhead{}& 
\colhead{}& \colhead{}& 
\colhead{} }
\startdata
SCDM	& 2.62--2.78 & 200 & 1818 & $9.0\times 10^{11}$ & .005 & .025 & .06& .02&$2.56 \pm 1.04$\nl
CHDM	& 2.57--2.73 & 200 & 1810 & $3.1\times 10^{11}$ & .035 & .070 & .37& .18&$4.30 \pm 0.77$\nl
\lcdm 	& 2.61--2.69 & 42  & 5178 & $7.3\times 10^{11}$ & .024 & .048 & .27& .10&$3.18 \pm 0.37$\nl 
OCDM	& 2.75--2.87 & 108 & 2873 & $6.9\times 10^{11}$ & .028 & .056 & .31& .13&$3.50 \pm 0.84$\nl
\tablecomments{
The redshift 
range covered by the pencils, the number of pixels, the comoving volume 
of a pixel, and the mass threshold,
chosen to reproduce the S98 number density per pixel, 
are given for each model.  Also listed are the fractions $p$ and $q$ of pixels 
with a fractional overdensity $\delg$ greater than 2.6 and 1.8, 
respectively (2.6 and 1.8 are $\delg$ for the two highest 
spikes in S98).  $P_1$ is the probability that at least one pixel out of 13 
will have 
$\delg$ greater than 2.6, and $P_2$ is the probability that at least 
one additional pixel will have $\delg$ greater than 1.8. The 
last column lists the 
best weighted fit to the
local biasing parameter as defined by equation (3) and discussed in Figure 6.}
\enddata
\end{deluxetable}	
}

We then multiply the count in each pixel by the inverse of the 
selection function at the center of the bin 
(normalized to unity at the maximum, $z=3.02$) 
and thus obtain a ``volume-limited" count in pixels, $N$.  We compute
the mean corrected count over the pixels, $\bar N=4.45$, and record
$\delg=(N-\bar N)/{\bar N}$ in each bin as the data for comparison
with theory.  This is the only operation we perform on the data; the
rest of the analysis is performed on the simulations.
Figure~\ref{fig:data}(b) shows the resulting distribution of $\delg$. 
This figure has two high isolated spikes, of $\delg=2.60$ and $\delg=1.82$.  
These spikes seem to be the most interesting features of these data.

The values of $\delg$ as reported by S98 are somewhat larger than the 
values quoted here 
because our definitions of the size of a pixel are somewhat different. 
Each of our pixels have the angular size of the S98 field and are the length 
of one their bins in redshift space.  Instead, S98 identify ``clusters''
(defined to be a group of galaxies whose proximity differs from a Poisson
distribution), and define the size of their spike by the edges of the
cluster that comprises it.
This definition of $\delg$ does not refer to a
well-defined scale and it therefore introduces unnecessary complications
in the comparison to theory. We therefore prefer to treat the counts
in pixels of a fixed volume.

\section{SIMULATIONS}
\label{sec:simulations}

\begin{figure*}[htb!]
\begin{center}
\resizebox{0.753\textwidth}{!}{\includegraphics{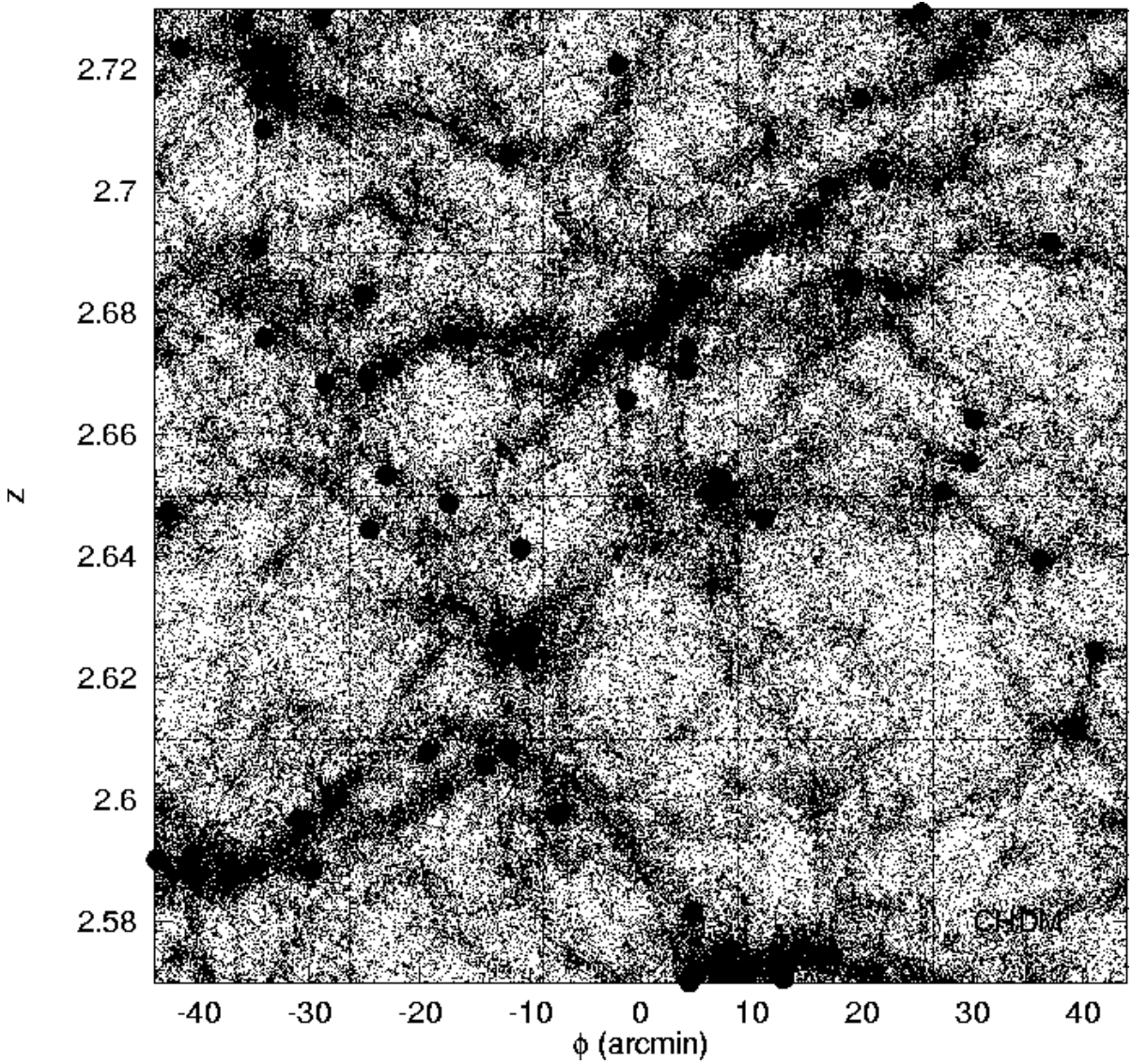}}
\end{center}
\caption{\capt
A slice in redshift space from CHDM at $z\sim 2.65$.
The simulation volume has a comoving side 75 \hmpc, and the thickness of 
the slice corresponds to an angle of 8.7 arcminutes, that of the 
Steidel \etal\ pencil.  The rectangular dotted lines indicate the angular 
width and the length in redshift space of the Steidel \etal\ pixels.
Shown are randomly sampled particles from the simulation (dots), 
and a 40\% random sample from the halos (solid circles),
which matches the observed number density of LBOs. The dark matter
distribution shows filamentary structure of moderate amplitude, but
several pixels show a high overdensity of halos, consistent with the
observed spikes in the distribution of LBOs.}
\label{fig:slice}
\end{figure*}

\begin{figure*}[htb!]
\begin{center}
\resizebox{0.857\textwidth}{!}{\includegraphics{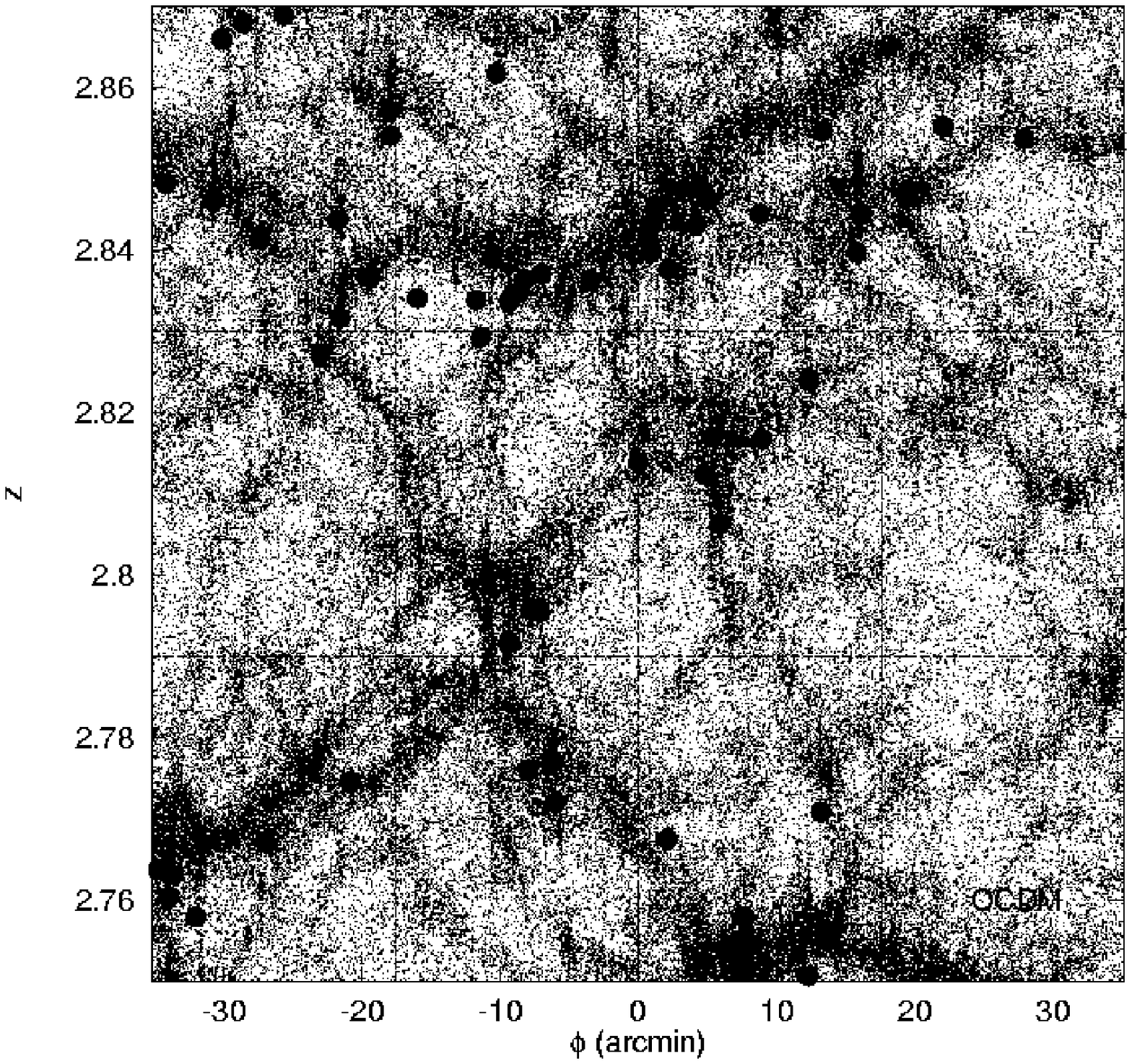}}
\end{center}
\contcaption{\capt The corresponding plot for OCDM at $z\sim 2.81$.}
\end{figure*}

\begin{figure*}[htb!]
\begin{center}
\resizebox{0.857\textwidth}{!}{\includegraphics{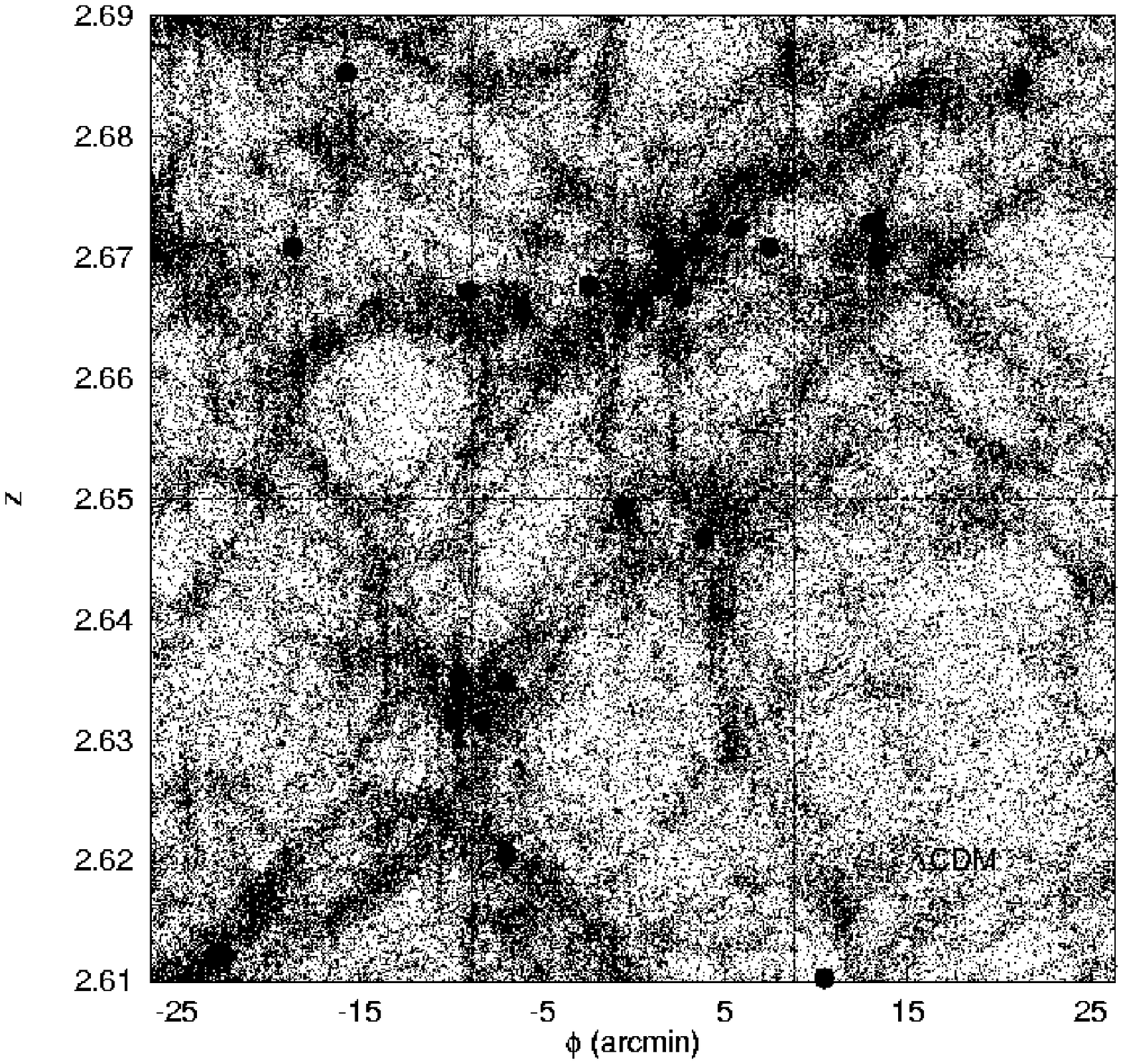}}
\end{center}
\contcaption{\capt The corresponding plot for \lcdm\  at
$z\sim 2.65$.  Because of the difference in geometry in 
\lcdm, the pixels are much bigger, and fewer pixels fit in the box.}
\end{figure*}

\begin{figure*}[htb!]
\begin{center}
\resizebox{0.857\textwidth}{!}{\includegraphics{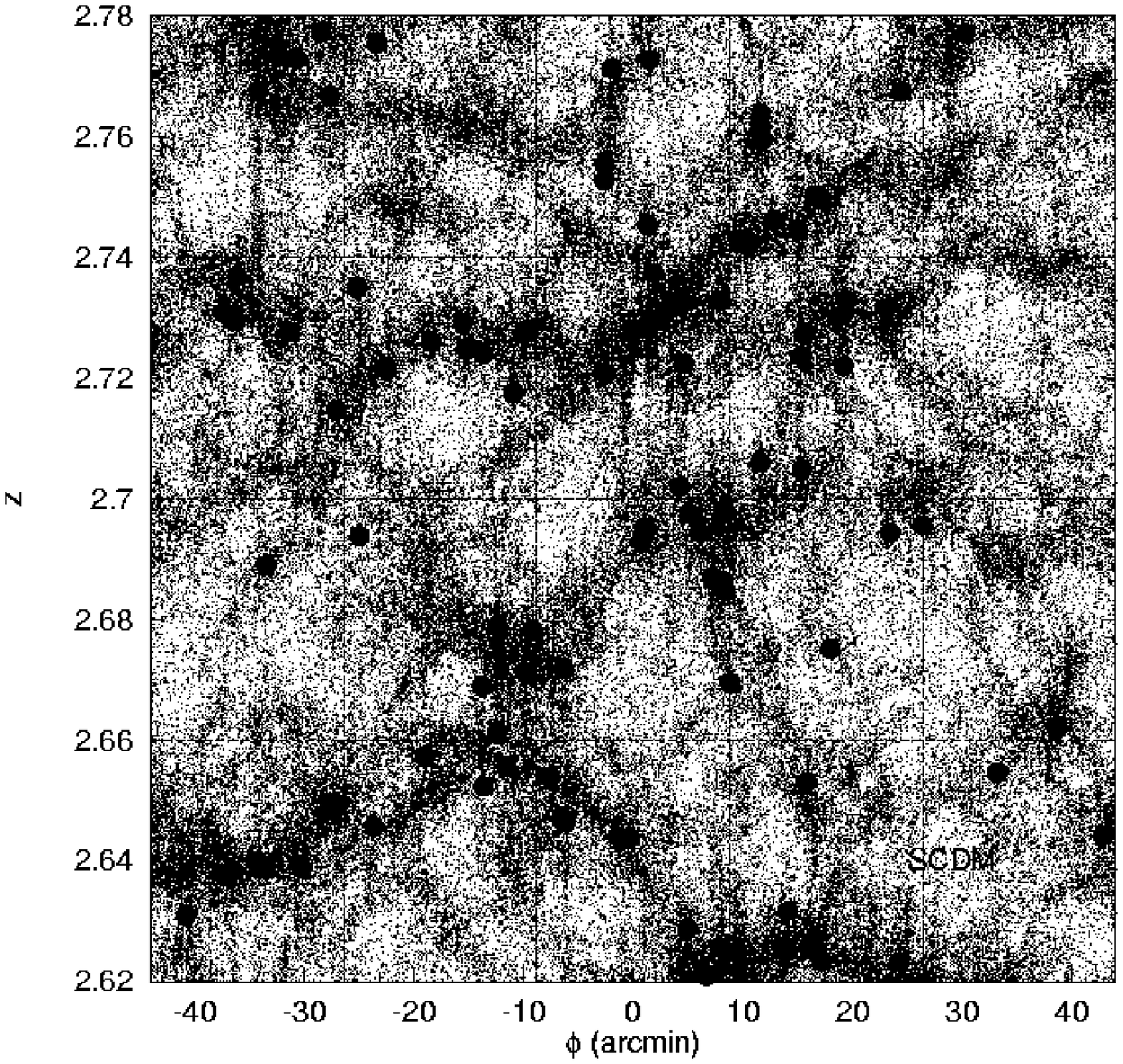}}
\end{center}
\contcaption{\capt The corresponding plot for SCDM at $z\sim 2.70$.}
\end{figure*}

We consider a suite of four different models for the formation of 
structure in the Universe:
\begin{enumerate}
\item  
One model is standard cold dark matter (SCDM),
with density parameter $\Omega=1$, Hubble parameter 
$h \equiv H_0/(100\,\kmsmpc)=     
0.5$, and a corresponding age of the universe today of $t_0=13$ Gyr. 
The fluctuation amplitude at 8 \hmpc\ is taken to be $\sigma_8=0.67$
at $z=0$, in order to approximate the abundance of Abell clusters. 
Unfortunately, this amplitude is far below the COBE normalization 
for this model.
\item 
Another model is the cold+hot dark matter model (CHDM) with 
$\Omega=1$, $h=0.5$ (thus $t_0=13$ Gyr),  
and $\sigma_8=0.72$. We consider two equal-mass neutrino
species contributing a total mass density of $\Omega_\nu =0.2$. 
This model is consistent with both cluster and COBE normalization.
It is termed CHDM-$2\nu$ in Gross \etal\ (1997a,b; hereafter G97a,b). 
\item 
Yet another model is the flat cold dark matter model with a nonzero
cosmological constant $\Lambda$ ($\Lambda$CDM) and $h=0.6$.  
Here, the mass density is $\Omega_0=0.4$ (relatively high, to obey the
constraints on the power spectrum from peculiar velocities in both
the Mark III and SFI catalogs,
Kolatt \& Dekel 1997; Zaroubi \etal\ 1997; Zehavi \etal\ 1998),
while the cosmological constant corresponds to $\Omega_\Lambda =0.6$. 
In this model $t_0=14.5$ Gyr.  
To simultaneously fit both the cluster abundance and the COBE data, we
include a slight tilt to the model, corresponding to a primordial fluctuation
spectral index $n_{\rm fluct}=0.90$, which gives $\sigma_8=0.88$. 
This model is called T$\Lambda$CDM in G97a,b.
\item 
Finally, we consider an open cold dark matter model (OCDM) with $h=0.6$
and $\Omega_0=0.5$ (a high value, again, for the same reason 
as for $\Lambda$CDM); for these parameters, $t_0=12.3$ Gyr.
Here, $\sigma_8=0.77$, consistent with both cluster and COBE normalization.
\end{enumerate}

We simulated the evolution of structure in these models in a purely
dissipationless manner. Our approach is based on 
a parallelized particle-mesh 
code, which we ran on the Cornell Theory Center SP2.  The code is described
in Gross (1997, Chapters 2--3); the simulations are discussed in G97a 
with further discussion of cluster abundance in G97b.  These simulations
include 57 million cold particles, with an additional 113 million hot
particles in the case of CHDM. The simulated box is 75 \hmpc\ in size, 
giving a mass per cold particle of about 
$2\times 10^9 \Omega_0$ \hmsun.
A single grid cell is 65 \hkpc\ wide in comoving coordinates,
corresponding to a physical width of 18 \hkpc\ at $z\approx 2.65$. 

The simulations were started at different redshifts, reflecting the
different fluctuation amplitudes of the various models at high redshift.  
This means that the earliest redshift at which halos were ``observed'' in the
simulations was somewhat different for each model.  The redshifts
that we analyze here are $z = 2.65$ for CHDM and \lcdm, $z = 2.70$ for
SCDM, and $z = 2.81$  for OCDM. 

We identify the halos in these simulations using the following procedure:
First, we use the density in grid cells to identify candidate halos at the
positions of local density maxima and neighboring cells 
with overdensities $\delta \rho /\rho > 50$.
Each candidate halo is then iteratively moved to the center of
mass of a sphere having a diameter equal to the grid-cell size of 65 $h^{-1}$
kpc. We define the mass of halos as the mass enclosed within a spherical region
whose density is sufficient for collapse and virialization (see Gross 1997,
Appendix C, for details).  At redshift $z=3$,
this corresponds to a $\delta \rho /\rho$ of 
178 for critical density models, 199 for \lcdm, and 203 for OCDM. 
Finally, we eliminate double counting by excluding smaller halos with 
centers inside larger halos.

As shown in G97a (Figure 10), we tested this approach by applying it to 
simulations run with lower spatial resolution and then comparing the 
resultant halo mass functions.  We found that for halo masses
used in this paper the grid size had an acceptably small effect.
The uncertainty in halo identification is even smaller at $z\sim 3$,
where the halos tend to be more isolated and contain less substructure.

We assume that one LBO resides in each massive halo.  
This assumption is motivated,
for example, by a semi-analytic model of galaxy
formation in which the brightest objects 
at $z\sim 3$ are predominantly 
starbursts in off-center collisions between sub-halo clumps
(Somerville, Faber \& Primack 1997; Trager \etal\ 1997; Somerville 1997; 
Somerville \& Primack 1998; Somerville, Primack \& Faber 1997).
In these simulations there is 
typically about one such starburst per massive halo at any given time.  
It would also be true in alternative
semi-analytic models in which the LBOs are identified as central 
halo galaxies (Baugh \etal\ 1997).

For this paper, we impose a minimum mass 
for the halos such that the mean number of halos in a pixel is 
2.5 times the number of redshifts obtained by S98 per pixel.  
We do this because S98 measured redshifts 
for only about 40\% of the candidates in their LBO survey. To
emulate the data, we therefore randomly select 40\% of our halos after setting
the mass threshold. This guarantees that each 
realization of the model matches the observed number of LBOs.  
The values of the mass threshold we find for each model are listed in Table 1.
They range from 
$M>3.2\times 10^{11}$ \msun\ for CHDM to $M>9.0\times 10^{11}$ \msun\ for SCDM.

Our procedure of matching the mean number density of LBOs has one 
important implication. Normally, one expects the abundance of halos 
of a given mass at a fixed redshift to be a strong function of the model. 
Some models have earlier galaxy formation than others. 
However, the exact relation between halos of a given 
mass and LBOs is not at all clear a priori, especially 
without the inclusion of gas dynamics and a reasonable model 
of star formation. Therefore, our elimination of this 
distinguishing factor allows us to concentrate instead on 
the variations in the density of objects --- especially 
the spikes.  

In Figure 2 we show slices of the underlying mass distribution and
a random sample of our identified halos for 
all four models. 
These slices have the angular thickness of the S98 pencil; the angular widths 
of the pencils are indicated in the figure. 
The comoving sizes of the pixels are different for each model 
because the fixed angular sizes and redshift interval
translate to different comoving scales in different cosmologies.
The pencils shown in Figure ~\ref{fig:slice} (a)
are only about 30\% as long as the observed one. 
We do not bother to select longer pencils (\eg, at angles to the box
sides) because we focus here on single-pixel
statistics.  
One might be concerned about the smaller redshift range represented by an
individual simulation volume, or by the difference in redshift between these
volumes and the observations, but as we will discuss in the next section,
the single-pixel statistics are not strongly dependent on these small changes 
in redshift.

The figure contains many very noticeable walls and filaments in
the underlying mass distribution, not unlike pictures of 
the galaxy distribution in slices at the current time in the Universe,
although the latter would also have massive clusters and larger voids 
than are visible here.  The most massive halos, indicated by filled 
circles, tend to lie within these sheets and filaments; this 
concentration of halos into relatively compact regions leads to spikes 
in the redshift distribution, as we discuss in \S~\ref{sec:stats}.

\section{STATISTICS OF SPIKES}
\label{sec:stats}

Using these simulations, we can investigate the statistics of clustering 
in the {\em observational plane.} In particular, we observe 
pencils in these simulations in redshift space, just as the observers do. 
The lengths of our pencils are constrained by the box size,
and they depend on the model that was simulated. 
Table 1 lists the redshift extent of the pencils in each model.
This corresponds to pencils of length $\Delta z=0.16$ (for CHDM and SCDM), 
$\Delta z=0.12$ (for OCDM), and $\Delta z=0.08$ (for \lcdm). 
We divide each pencil into 2 to 4 pixels, 		    
each identical to the observed pixels. This yields a total of 
200 pixels for CHDM and SCDM, 108 for OCDM, and 42 for \lcdm. 
Another reason for variations in the number of pixels that fit in the box 
is the fact that the fixed angular size of the observed pencils 
corresponds to different comoving sizes in the different models. 

\begin{figure*}[htb!]
\resizebox{\textwidth}{!}{\rotatebox{-90}{\includegraphics{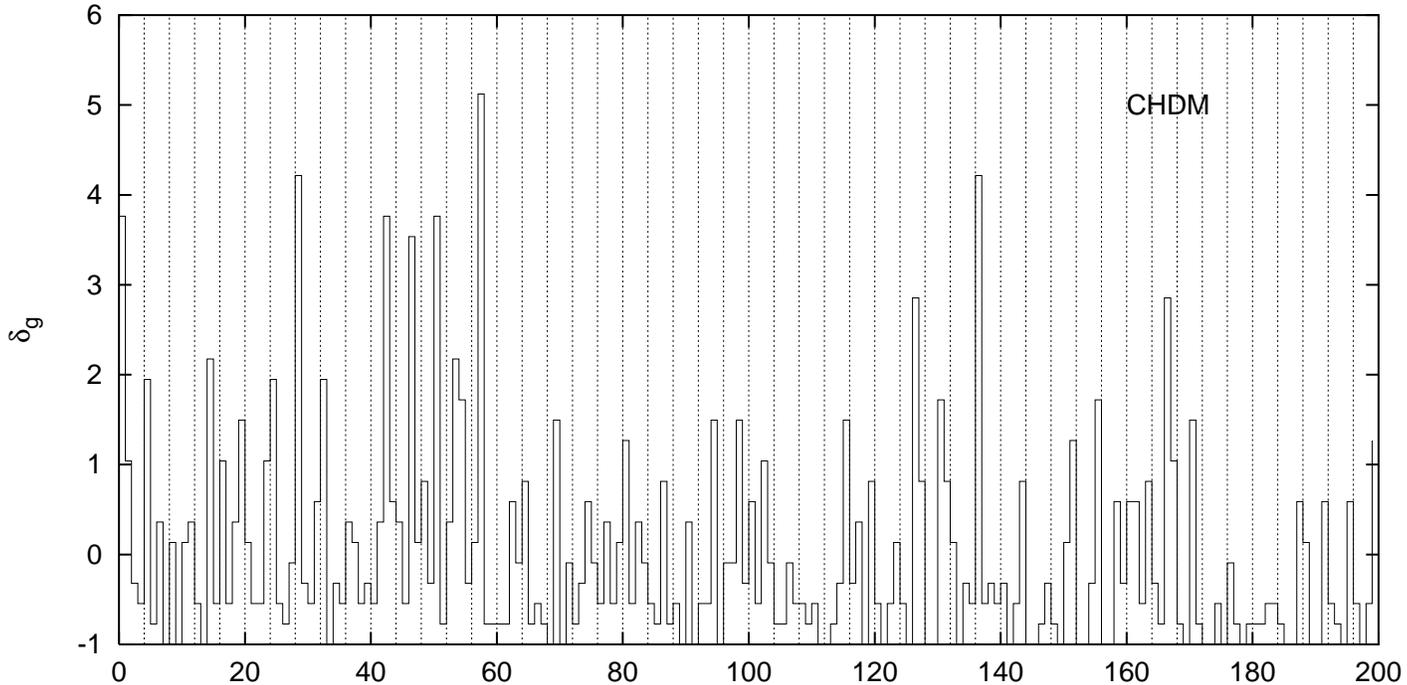}}}
\caption{\capt
Fifty pencils (containing 200 independent pixels) laid
end to end for the CHDM simulations. Each of the independent pencils
is separated by a vertical dotted line, and the 
pencils are arranged by increasing $\theta$ in groups of increasing $\phi$.
The plotted quantity, $\delg$, for each
pixel is the fractional excess of galaxies within each pixel over the
average in the realization.}
\label{fig:spikes}
\end{figure*}
In Figure~\ref{fig:spikes}, we show 
$\delg$ 
in the pixels of 50 independent pencils 
from one random sampling of CHDM stuck together in a row. Each
pencil has a length of four pixels, and we separate the
pencils with vertical dotted lines. 
We can afford to deal with pencils that are only four
pixels long, because we do not intend to investigate here 
the effects of correlations between the pixels along a pencil. 
In particular, we will focus on the question of the probability of finding a
single pixel with an overdensity as high as that observed by S98. 
Since both the observations and the simulations (Figure~\ref{fig:spikes}) 
suggest that the highest peaks are only one pixel wide, this indicates 
that single pixel statistics may provide 
a sensitive test for how probable such pencils are in various cosmological 
models.  

\begin{figure*}[htb!]
\begin{center}
\resizebox{0.48\textwidth}{!}{\rotatebox{-90}{\includegraphics{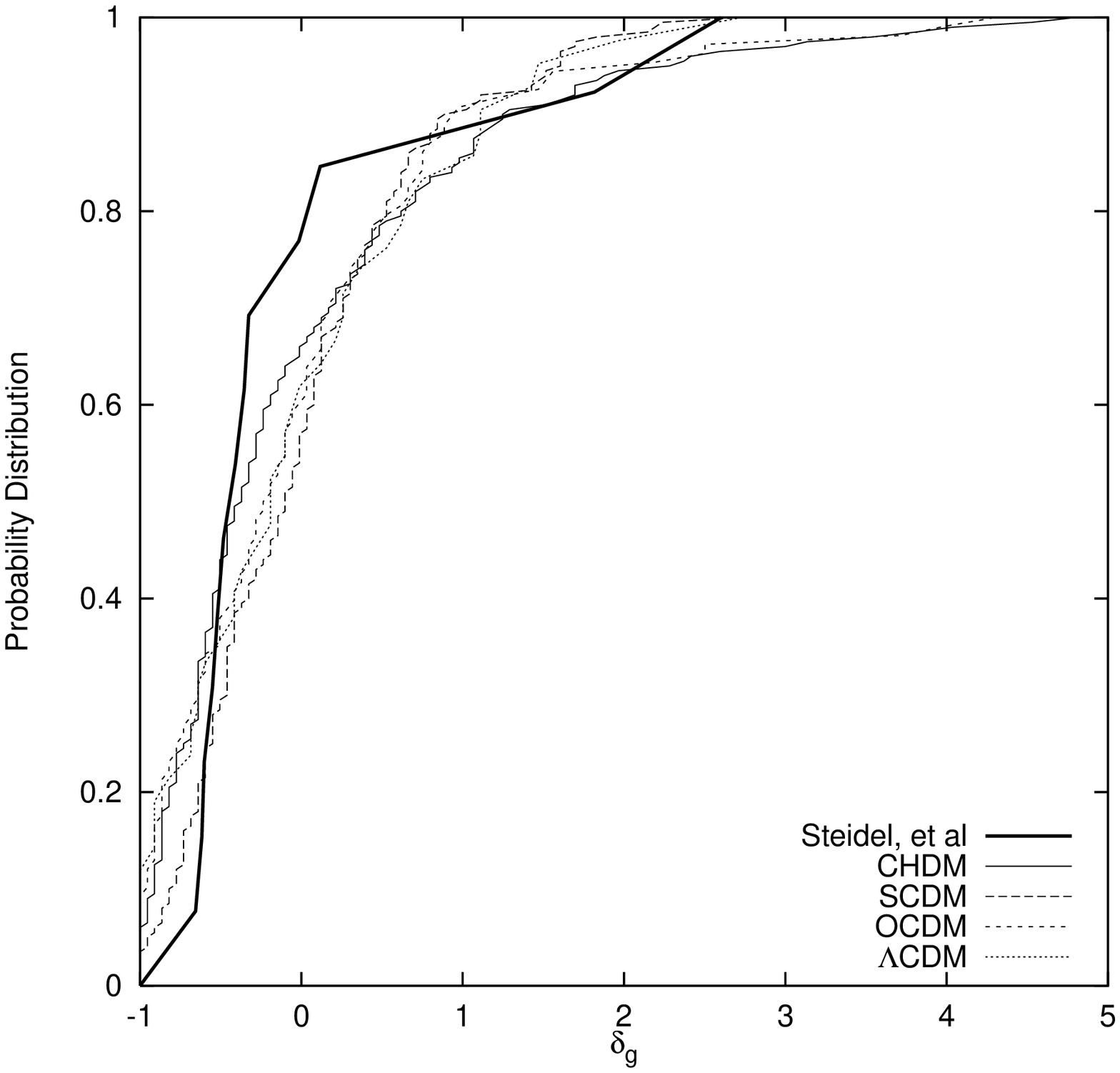}}}\hfill
\resizebox{0.48\textwidth}{!}{\rotatebox{-90}{\includegraphics{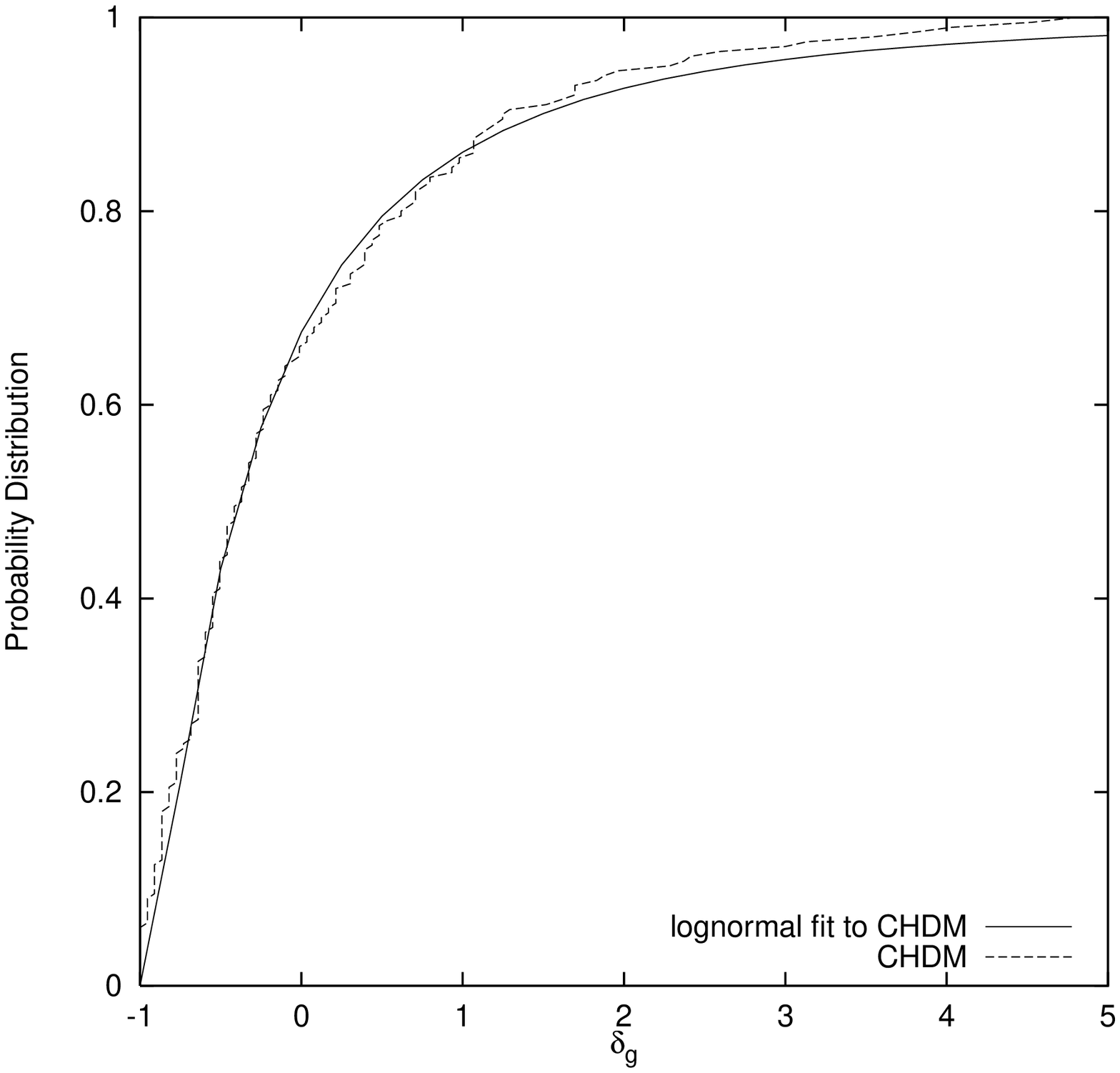}}}
\end{center}
\caption{\capt
{\em Left panel:} A plot of the cumulative distribution of
$\delg$ for the Steidel \etal\ (1997) data as well as for the
halo distribution in the four models considered here. The distribution
reflects the relative excess of halos in each pixel.
{\em Right panel:} The
cumulative distribution for the CHDM model compared with the best
lognormal fit to that distribution.}
\label{fig:dist}
\end{figure*}

For each simulation, 
we randomly sample 40\% of the halos above the threshold
five times for each model. We
then average over these samplings in order to examine the distribution and
statistics of the number of halos in a pixel. In Figure~\ref{fig:dist} 
(a) we plot the cumulative distribution of the relative excess in the number 
of galaxies per pixel, 
$\delg$, for each of the four models as well as for the data shown in
Figure~\ref{fig:data}(b). 
We also plot, in Figure~\ref{fig:dist}(b), the cumulative distribution of 
$\delg$ for one of the models, CHDM, 
compared to the best lognormal fit.

In the CHDM model, the highest pixel has $\delg=4.8$, almost
twice as large as the $\delg=2.6$ of the highest observed pixel.  
In fact, there are several pixels in the model with galaxy density higher 
than the highest observed pixel.
The cumulative distribution shows that the probability that the galaxy 
density contrast in a randomly-chosen pixel of this model
will exceed $\delg=2.6$ is $p=0.035$.  The probability
$p$ of exceeding this value is given, for each of the models, in Table 1.

If instead of observing one single pixel, one observes $n$ independent
pixels, then the probability $P_1(n)$ that at least one of the pixels
will exceed the threshold $\delg=2.6$ follows from the
binomial distribution:
\begin{equation}
P_1(n) = 1-(1-p)^n \; .
\label{qqq1}
\end{equation}
Getting at least one such peak is just one minus the probability of no
such peak. Note that equation~(\ref{qqq1}) assumes that there are no
correlations among the observed pixels.

In the absence of correlations among the pixels, one could further ask what is
the probability $P_2(n)$ that one pixel will exceed $\delg=2.6$ while
at least one additional pixel will exceed a certain lower threshold
of probability $q$ per pixel.
The second-highest pixel in the S98 data has $\delg=1.8$.  
The probability $q$ of exceeding this lower threshold for each of the models
is also given in Table 1. 
The probability $P_2(n)$ encompasses all possibilities except 
(a) no pixel above the highest threshold or 
(b) exactly one pixel above the highest threshold with no additional 
pixel exceeding the lower threshold. This probability therefore follows 
directly from the multinomial distribution and is
\begin{equation}
P_2(n) = 1-(1-p)^n -np(1-q)^{n-1} \; . 
\label{qqq2}
\end{equation}

For $n=13$ randomly chosen pixels, the probability that there will be
at least one above $\delg=2.6$ is 37\% for CHDM. The
probability of at least one pixel above $\delg=2.6$ 
and at least one additional pixel above $\delg = 1.8$ is 18\%.  
These numbers follow from the averaged data for 5 random CHDM selections.  
In some of the individual selections, the probabilities were as high as 
$50\%$ and $28\%$.  
If, instead of randomly sampling 40\% of the halos, we set 
the threshold higher and use all of the halos, 
$p$ and $q$ are 0.05 and 0.075, which gives $P_1 =$ 49\% and 
$P_2 =$ 23\%.
The other model simulations have slightly lower probabilities,
as listed in Table 1.

Probably the biggest surprise here is that all the models except SCDM 
predict that a large spike at $z \sim 3$ is not a very unusual event.
All models except SCDM yield a peak as large as the highest observed 
by S98 at least 27\% of the time, while 
such a peak occurs in SCDM in 6\% of such pencils. 
It is clear from these statistics that SCDM has less clustering at 
these high redshifts.  
It may be that this is due to the shape of the power spectrum
$P(k)$, which for SCDM has a shallower slope 
on the large-$k$ side of the peak than any of the other models.  
In SCDM the ratio of small-scale to large-scale power is 
higher than in other models, which results in a more even 
distribution of galaxies, with more galaxies in the voids (for a
visualization showing this, see Brodbeck \etal\ 1998).  The difference 
in how the LBOs are distributed in the models can be seen clearly 
by comparing the CHDM and SCDM pixels of Figure 2.
In the Appendix, we discuss an analytic formalism that helps to understand the
dependence of the spike probability on the shape of the power spectrum.
One could further test whether the lack of clustering in SCDM is due 
to the slope of the power spectrum by looking at simulations of an 
$\Omega=1$ CDM model with a steeper power spectrum (one 
which has more large scale power for the same $\sigma_8$ 
normalization), e.g., $\tau CDM$ (see Jenkins \etal\ 1997). 

The model distributions shown in Figure~\ref{fig:dist} are remarkably similar. 
One might have thought that low-$\Omega$ models 
have earlier structure formation and therefore are more likely to show a 
big spike compared to Einstein-deSitter models with $\Omega =1$. 
However, there is a competing effect --- the fixed angular size of the 
pixels at a given redshift and the fixed redshift interval
mean that the comoving volume of a pixel is larger in open models. 
The density contrast quoted thus refers to a larger scale, and is 
therefore expected to be lower. 
The comoving volume of a pixel is given, for each of the models, in Table 1.

We do not anticipate that there is a significant problem with the fact 
that we sample at $z \sim 2.65$ rather than the observed $z \sim 3$. 
In linear theory, the growth of the mass density fluctuations themselves
between these two epochs is rather small, less than 10\%, and the evolution
of the {\it galaxy} density fluctuations is expected to be much weaker.  
In order to demonstrate that, 
we analyze the CHDM simulation at $z=1$ as if it had the geometry of the 
simulation at $z=2.65$ (i.e., we take the comoving positions at $z=1$, 
assume all objects had the same comoving positions at $z=2.65$, and at this
earlier time put down pixels of the S98 size); this has the effect of 
including any evolution in the amplitude of the fluctuations while 
ignoring changes in geometry.  
We find that the effect on the statistics is very small:
$P_1$ = 37\% (unchanged from $z=2.65$), and $P_2$ = 11\% (slightly lower than 
for $z=2.65$).  
Thus the main effect in going from lower to higher $z$
is how the pixel geometry is affected, but this effect is rather small
in going from $z=2.65$ to 3. 
For example, if the CHDM simulation (at $z=2.65$)
is analyzed as if it were at $z=3.0$, the probabilities become
$p=0.042$ (vs. 0.035), and $q=0.083$ (vs. 0.070), which give $P_1$ = 43\%
and $P_2$ = 23\%.
Thus, the likelihood of the spikes in each of the models would be,
if anything, slightly increased, had the analysis been done at $z = 3$.
The general effect we are seeing here is that the massive
halos trace the large scale structure and the clustering of
these halos evolves rather slowly, while the dark matter
meanwhile becomes steadily more clustered -- thus the bias of
the $10^{12} M_\odot$ halos compared to the dark matter will
be larger at high redshift.

We have done the same analysis for SCDM at $z=1$.  In this case, evolving in
redshift from $z=2.70$ to $z=0.93$ (changing the scale factor by a factor
of 1.9) is equivalent to changing the amplitude of the model by a factor of
1.9, which is then close to the COBE normalization for SCDM.  Again, the
effect of this change on the shape of the probability distribution and on the
statistics is small: for this case, $P_1 = 0.06$ and $P_2 = 0.01$.
One should note that in this analysis and that described in the previous
paragraph, we still require that the mean number of LBOs per pixel is the same
in each calculation, and set the mass threshold accordingly.  This cancels
the dominant effects
\linebreak
\begin{center}
\resizebox{0.47\textwidth}{!}{\includegraphics{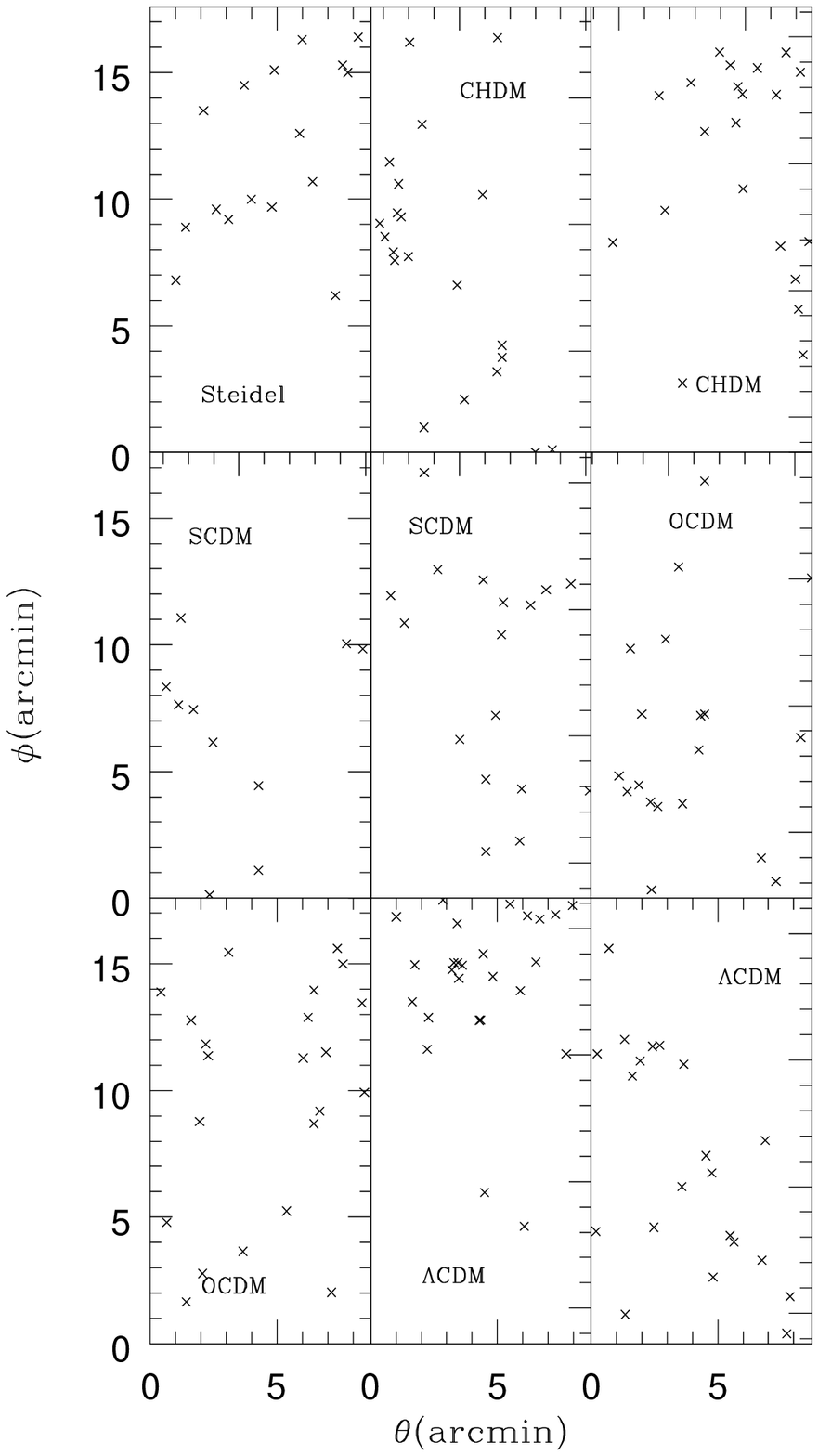}}
\end{center}
{\capt\figcaption{\capt The angular distribution of halos within high spikes in
redshift. The data from the highest spike of Steidel \etal\ (1997) are 
shown in panel a, while
the other eight panels show two examples of high spikes from each of
the simulations. The angular size of the panels correspond to the
angular size of the pencil survey of LBOs.
\label{fig:angular}
}}
\vskip 2ex
that would have been seen from changes in the evolution or
the normalization.  Since the normalizations of the four different models at
$z=3$ are not different by more than this factor of $\sim2$, we can
conclude that any difference in the spike probabilities
between the models are not primarily due to 
differences in their normalizations.
The dependence of the spike probabilities on normalization is discussed 
further in the Appendix. 

One other notable feature of Figure~\ref{fig:dist} is the fact that,
in each of the simulations, between five and ten percent of the pixels
are empty, while there are no empty pixels in the data of 
Figure~\ref{fig:data}.
However, the fact that pencils in other surveyed regions do contain 
empty pixels (Adelberger, \etal\ 1998) suggests that this
does not represent a significant disagreement between theory and observation.

The spatial distribution of halos on the plane of the sky in each of the 
models is qualitatively similar to that observed.
Figure \ref{fig:angular}
shows the distribution on the sky of the halos in two 
``spiky'' redshift bins for each model, compared to the data
for the highest spike.  We can see
that halos in the spikes are typically part of a filament or a wall stretching
across the pixel. 
Thus, the spikes in redshift do not
correspond to extreme localizations in space, as would be the case for 
today's clusters; rather the structures
are only slightly localized in angle, and typically only in one dimension.
But these ``spikes'' do end up in rich clusters at the present epoch.

\section{BIASING}
\label{sec:biasing}

Another quantity that we can easily measure in our suite of simulations is the
biasing relation between the halo and mass-density fluctuations.
This will directly connect to the interpretation by S98 of their own
observations. 
In each pixel of the simulations, we have already calculated the density 
of galactic halos, $\delg$, and we now compare it to the overdensity of 
the background matter, $\delm$, as a direct measure of the biasing on 
the scale of the pixels. 
This biasing refers to the specific definition of halos specified in
\S\ref{sec:simulations}, and is particularly dependent on the choice of the
halo ``edge,'' which for our purposes is the radius which encloses a mean
density sufficient for collapse by $z\sim 3$. 

\begin{figure*}[htb!]
\resizebox{\textwidth}{!}{\includegraphics{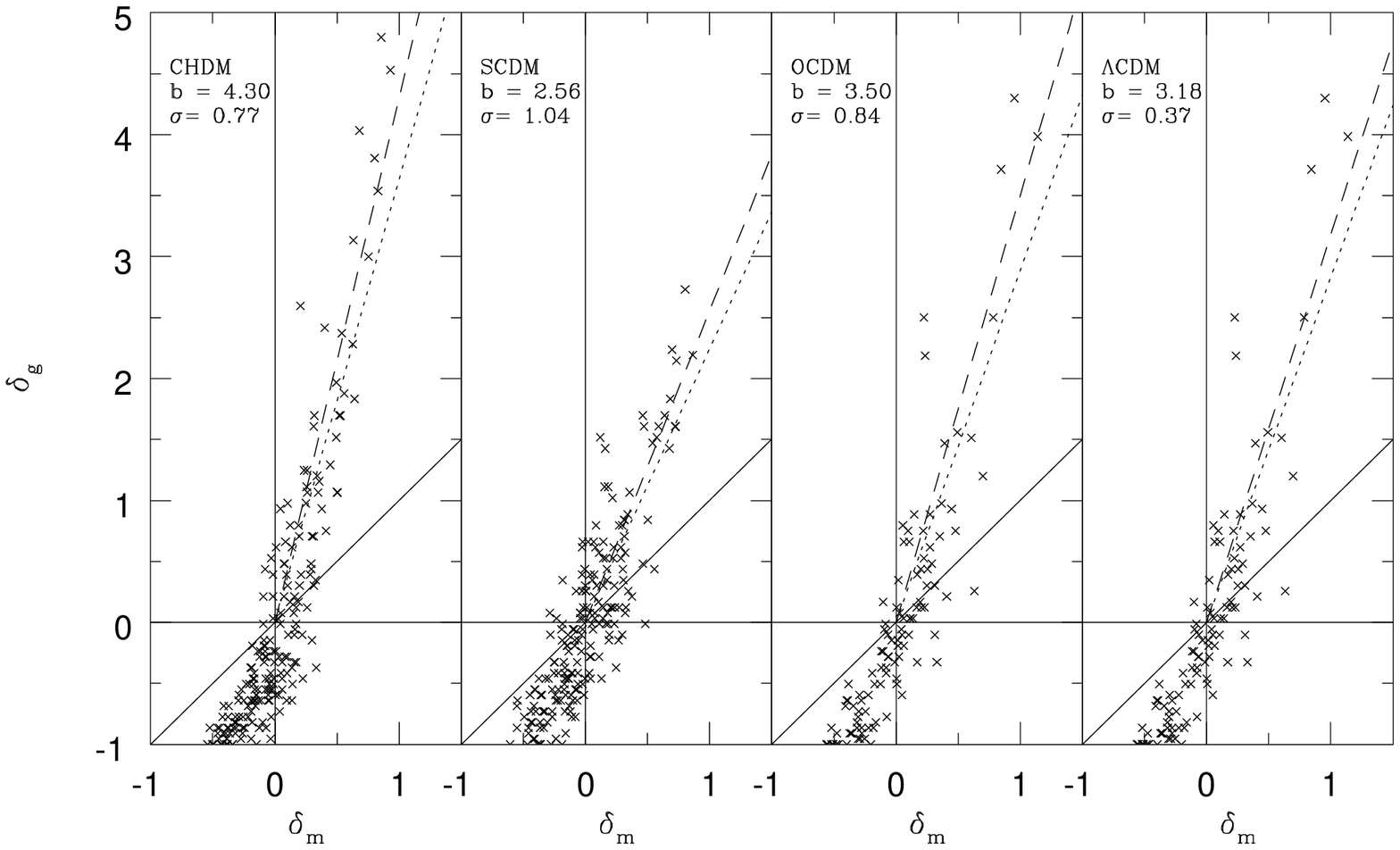}}
\caption{\capt
Plots of galaxy overdensity in a pixel $\delg$
versus the dark matter overdensity in the same pixel $\delm$
for each of the simulated models considered here. The solid line in
each curve corresponds to biasing parameter $b=1$. Ignoring underdense
regions ($\delm < 0$), we also plot
two regression curves to these data. The dotted curve corresponds to a
least squares fit, while the dashed curve represents the best fitting
straight line when we weight each pixel using the Poisson errors due to
the finite number of halos. The values of $b$ and $\sigma$ quoted in
the figure correspond to the best slope and its error using the weighted fit.}
\label{fig:biasing}
\end{figure*}
In Figure~\ref{fig:biasing}, we plot for each of the simulations,
at the corresponding redshift near $z\sim 3$ (\S~\ref{sec:simulations}),
the galaxy overdensity in a pixel versus the dark matter overdensity in 
the same pixel.  For each pixel, the local biasing is just
the ratio of densities 
\begin{equation}
b_{\rm local}= \delg/\delm.
\end{equation}

To get a measure of the mean biasing of the entire sample at that epoch, we
calculate the regression of $\delg$ upon $\delm$.
Since we are interested primarily in regions that will
ultimately collapse, and since the biasing relation is not linear at
$\delm <0$, we use only the positive part of the figures
($\delm > 0$) to calculate the regression. 
This approach yields the dotted lines. A 
different approach is to find the best fitting straight line while taking into
account the errors (due to Poisson counting) in $\delg$ within
each pixel. This weights the highest $\delg$ peaks more strongly and leads to
the dashed lines in each panel of Figure~\ref{fig:biasing}.

Using the weighted measure of mean biasing, we find that the mean
biasing parameter $b$ varies from a high of $4.30\pm 1.14$ for CHDM,
to a low of $b=2.56 \pm 0.52$ for SCDM (Table 1). 
The error bars we quote correspond to the formal error in the slope fitting. 
Hence, $b$ is constrained to the range $b\simeq 2-5$ for all models. 
The low-$\Omega$ models appear less biased than CHDM with $\Omega_{\rm m} =1$,
but recall that the volume of a pixel in the low-$\Omega$ models is
larger and therefore the biasing measured here refers to a larger scale.

Average biasing does not tell the whole story since there is
considerable dispersion in the local biasing parameters,
and the selection of high peaks of $\delg$ clearly biases the local
biasing parameter to larger values. For example, in 
both CHDM and OCDM there are pixels having local biasing 
greater than 10. 
In CHDM, most of the regions corresponding to high spikes in the data have 
a bias of about six, while in the other three models the bias in the regions 
with high spikes ranges from 3.5-4.5.
These values for the bias are similar to the values that S98
found were necessary to get a reasonable probability of large spikes.
It was not necessary in our analysis to make all of 
the assumptions that were necessary for S98 to do the 
analysis in the theoretical plane, however, which in particular includes
the assumption of quasi-linear biasing.  
Our own simple analytic treatment of the dependence of bias on the shape and
normalization of the power spectrum are discussed in the Appendix.

\section{CORRELATIONS}
\label{sec:correlations}

Since the probability of a spike in the distribution of LBOs as observed
by S98 is not too small for any model considered here
(except possibly SCDM),
it does not represent a good discriminator between such models.  
In this comparison, we had to normalize out what might otherwise be
the dominant effect --- the exponential dependence of the number 
density of objects on the power-spectrum normalization at high redshift
--- because it depends on a more accurate identification of halos as LBOs
than we were willing to deal with in the present work.  
Still, there might be hope for better discrimination between models using a 
statistic that is insensitive to the number density of objects, 
such as the autocorrelation function.

\begin{figure*}[htb!]
\begin{center}
\resizebox{0.48\textwidth}{!}{\rotatebox{-90}{\includegraphics{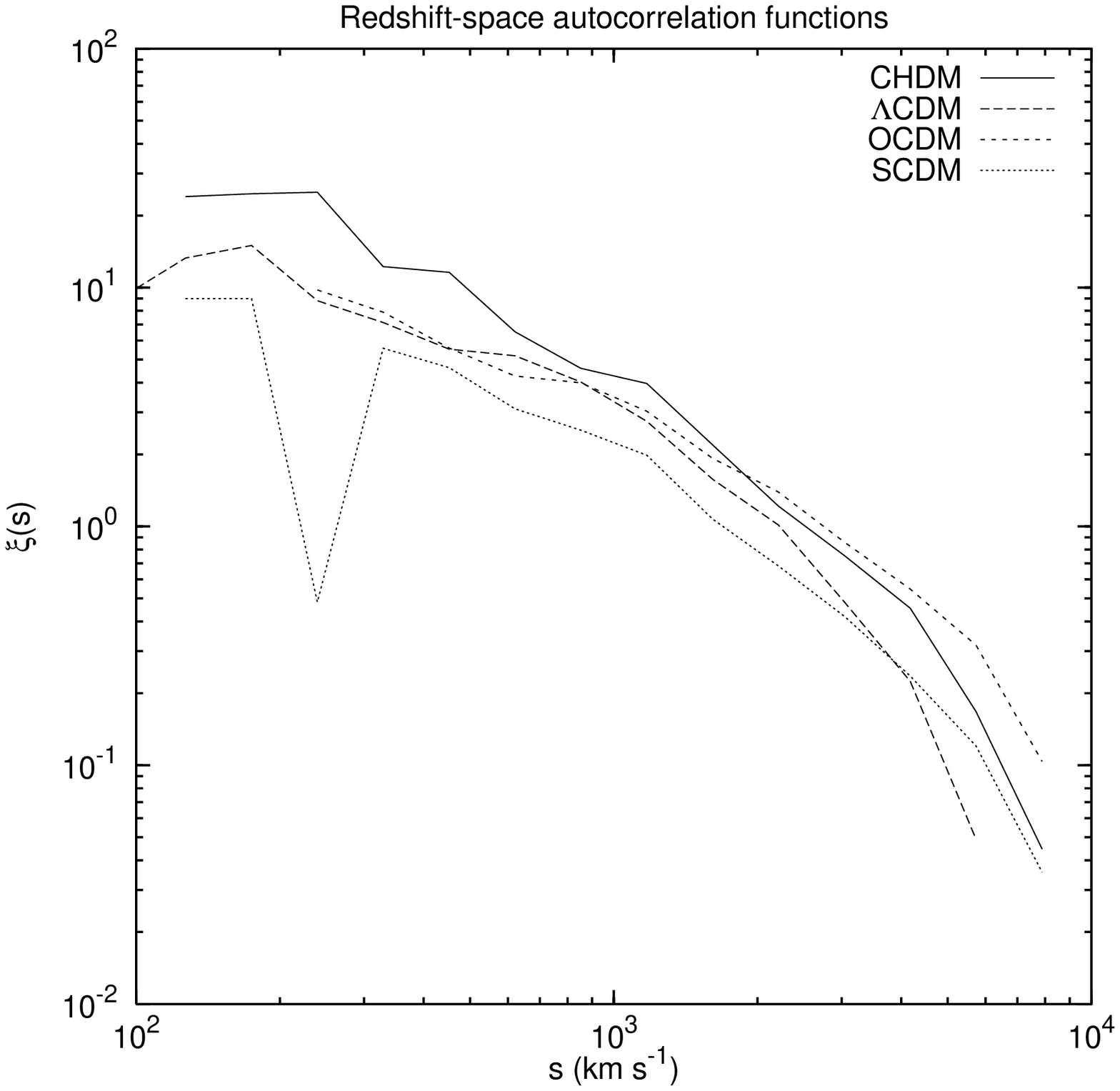}}}
\resizebox{0.48\textwidth}{!}{\rotatebox{-90}{\includegraphics{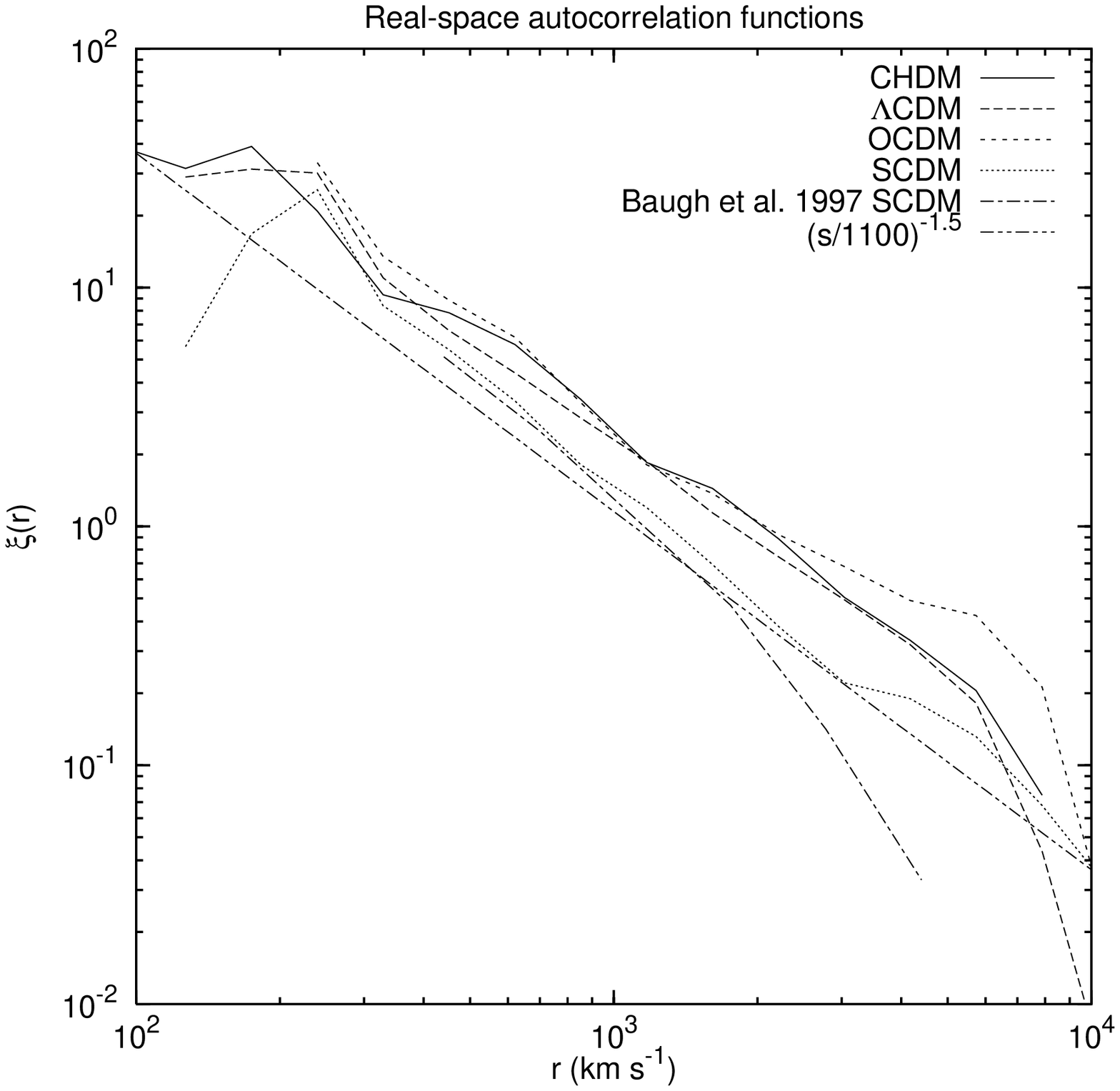}}}
\end{center}
\caption{\capt
Redshift space ({\em left}) and real space ({\em right})
autocorrelation 
functions for each model. Functions are calculated in a 
volume corresponding to all the ``pencils'' described in 
\S\ref{sec:stats} for a given model, left in their original 
positions in the computational box.  In calculating the redshift space
correlations, each object has had its                       
peculiar comoving velocity along the line of sight added 
to its coordinate distance, to obtain the observed redshift.  
The real space data has no such correction, and the dot-dashed line 
is a comparison to the semi-analytic results for SCDM from 
Baugh \etal\ (1997).
Our correlation length is very similar to theirs, but our slope seems to
be shallower at large separations.
Note that the conversion from km s$^{-1}$ to comoving $h^{-1}$ Mpc is not
merely the Hubble parameter because the distance is measured between two
distant objects rather than one distant object and the observer.  The correct
conversion factor, required to preserve angular separations, is $cz/d$, where
$d$ is the coordinate distance to redshift $z$,
corresponding to 278$h$ km s$^{-1}$ Mpc$^{-1}$ for SCDM. 
Thus the correlation length $r_{0}$, such that 
$\xi(r_{0}) = 1$, for the  Baugh \etal\ (1997) curve is 
$r_{0}$ = 4.0 $h^{-1}$ Mpc, corresponding to $1.1
\times 10^{3}$ km s$^{-1}$.}
\label{fig:corrfn}
\end{figure*}
Figure~\ref{fig:corrfn} shows the three-dimensional redshift-space
autocorrelation functions,
and also the corresponding real-space autocorrelation functions. 
In both cases,
separations between pairs of LBOs were estimated using
the redshift $cz$ as the distance from the observer.  Since such a
definition must preserve angular separations, the relation between
redshift-determined separations $s$ and comoving separations $r$ is
\begin{equation}
s = r \frac{cz}{d(z)},
\end{equation}
where $d(z)$ is the coordinate distance to redshift $z$.
The ratio $cz/d$ reduces to the Hubble parameter if $z\ll 1$, but for the
high redshifts listed in Table 1, it is 281, 278, 206 and 256
$h$ km s$^{-1}$ Mpc$^{-1}$ for CHDM, SCDM, $\Lambda$CDM and OCDM, respectively.

In calculating the redshift-space correlation function, the 
velocity along the line of sight is added in a vector 
fashion.  The correlation function is then calculated in the 
usual manner, by creating a randomly distributed catalog ten 
times larger than the halo catalog and dividing the number 
of halo-halo pairs by a tenth of the number of halo-random 
pairs, as a function of comoving distance in km s$^{-1}$. 
The real-space result is consistent with the semi-analytic estimation 
of Baugh \etal\ (1997), with the same correlation length of 
$\sim$4 $h^{-1}$ Mpc for the SCDM case, but with a flatter slope
in our simulations.
In Table 2, we give the best fit values of $r_0$ 
(in $\hmpc$) and $\gamma$ for each of the models, in both real and 
redshift space, for $\xi (r) = (r/r_0)^{-\gamma}$ fit for $r\leq 5$ $\hmpc$.

{\footnotesize
\begin{deluxetable}{ccccc}
\tablenum{2}
\tablecolumns{5}
\tablecaption{}
\tablehead{
\colhead{}              &
\multicolumn{2}{c}{Real Space} & \multicolumn{2}{c}{Redshift Space} \\
\cline{2-3} \cline{4-5}
\colhead{Model}         &
\colhead{$r_0$}         & \colhead{$\gamma$} &
\colhead{$r_0$}         & \colhead{$\gamma$}\\
\colhead{}              &
\colhead{($\hmpc$)}     & \colhead{} &
\colhead{($\hmpc$)}     & \colhead{}}
\startdata
SCDM &  3.19&   1.69& 3.27& 1.68 \nl
CHDM &  4.97&   1.55& 5.12& 1.59 \nl
\lcdm&  6.47&   1.48& 7.27& 1.49 \nl
OCDM &  4.72&   1.52& 5.01& 1.59 \nl
\enddata
\tablecomments{The best fit value of $r_0$ and $\gamma$ are given here for
each model.  If we set $\gamma=1.8$, to agree with local observations, the
values for $r_0$ are slightly but not significantly lower.}
\end{deluxetable}
}

Interestingly, our correlation length for SCDM at $z\sim3$ is
comparable in comoving coordinates to that of galaxies today, while
the other models have somewhat larger correlation lengths; the
logarithmic slopes are closer to $\gamma=-1.4$ than the $\gamma
\approx -1.8$ observed in galaxy redshift surveys today.  Their shallower
$\gamma$ may suggest that the LBOs are distributed more like sheets
than filaments at $z\sim 3$, and and that they evolve into more
filamentary structures with time --- or perhaps that they do not
evolve into typical bright galaxies at the present epoch.  The fact
that SCDM has a lower correlation length than the other models perhaps
just reflects the well known fact that SCDM, with its power spectrum
$P(k)$ having a broad peak, has a matter correlation that becomes
negative at smaller separations than the other models we consider,
which have $P(k)$ falling faster on the large-$k$ side of the peak
(see e.g. Holtzman \& Primack 1993 for a discussion of this for the
same cosmological models, in the context of the cluster
autocorrelation function).  Small differences in high-redshift
correlation lengths and slopes might help discriminate
between models, once enough observational data has been collected to
represent a ``fair sample'' of the universe.

\section{CONCLUSIONS}
\label{sec:conc}

We find that large peaks (``spikes'') in the observed redshift distribution of
LBOs within very deep pencils at $z\sim 3$ are a common
occurrence among the competing models for the formation of large-scale 
structure, when LBOs are identified as massive dark matter halos in our 
high-resolution simulations. 
Spikes of the sort observed by Steidel \etal\ (1998) 
occur frequently in simulations of cold+hot dark matter, open cold 
dark matter, and a model of CDM containing a cosmological constant,
and occasionally in standard cold dark matter.

Note that although our SCDM model has $\sigma_8=0.67$ (cluster normalization), 
increasing the amplitude by a factor of $\sim 2$ (to near-COBE normalization) 
hardly affected the spike probabilities (as discussed in \S\ref{sec:stats}). 
The fact that such spikes are expected in these models means that the 
existence of one or more large spikes cannot discriminate among competing 
models, although additional statistics may begin to do so against SCDM. 
We speculate that the lower spike probabilities in SCDM, and the fact that 
SCDM is the model with the least biasing, 
are a consequence of the shallower slope of its power spectrum.
Of course, we have chosen enough halos in all of our simulations to guarantee 
that the mean number of LBOs at $z\sim 3$ agrees with the observations.
More realistic treatment of galaxy formation (or LBOs) may 
yield different abundances of galaxies at high redshift, and consequently,
differing likelihood of large spikes.

We also found that our models give a mean biasing parameter in the range 
$b\sim 2-5$.  This refers to scales of order 10 \hmpc\ at $z\sim 3$,
and to a definition of $b$ via unweighted linear regression at $\delta>0$.
High-density regions have somewhat higher local biasing values, 
typically around six for CHDM.
In some pixels the local biasing parameter is even greater than 10.
Recall that our definition of $b=\delg/\delm$ refers to pixels of fixed size 
while S98 refer to a smaller volume. 
Note also that our analysis is done in the ``observational plane'' 
and it involves the fully non-linear fluctuations in the numbers of 
LBOs ($\delg$) and matter ($\delm$),
compared to the analysis in the ``theoretical plane" of S98.
We find little or no dependence on $\Omega_0$ or $\Lambda$,
but this is partly due to the differences in pixel volumes with 
different cosmological parameters. 
However, our cold plus hot $\Omega=1$
model has the highest mean bias, $\bar{b}=4.3$.
It should also be noted that our low-$\Omega_0$ models have higher values of 
$\Omega_0$ than the ones that S98 consider, which may be one reason 
why there is less difference between our models.
As discussed in Section 4, the clustering of LBOs (as 
described by $\delg$) is not strongly evolving, while the dark 
matter ($\delm$) clusters more strongly at low redshift, 
resulting in a decrease in the bias.

The observation of high spikes is thus consistent with standard
cosmology and with straightforward statistical correspondence of LBOs
with the most massive dark-matter halos at $z\sim 3$.
S98 say in their abstract that ``in a cold dark matter 
scenario the large bias values suggest that individual Lyman-break 
galaxies are associated with dark halos of mass $M\sim 10^{12}$ \msun, 
reinforcing the interpretation of these objects as the progenitors of 
massive galaxies at the present epoch." 
Our results imply that the LBOs do have the clustering properties
of massive dark matter halos, 
but it is important to note that this does not {\em necessarily} imply 
that they are the progenitors of present-day massive spheriods.  
These results are also consistent with a model like that proposed
by Somerville, Primack, \& Faber (1998) in which although most of the LBOs are 
found in halos of mass $M\sim 10^{12}$ \msun, the LBOs themselves may be small 
star-bursting satellites of a central massive object.

The halos identified with LBOs at $z\sim3$ 
are distributed on the sky much like those in the highest spike 
of S98 (see Figure 5), with no evidence of a central concentration.  But
if one follows the evolution of the regions with the highest spikes, 
virtually all of them become massive clusters 
(M$ \gtrsim 3\times10^{14}$ \msun) at the present time.
In our CHDM simulation, for example, those halos that correspond to Abell 
richness $\geq 0$ clusters at $z=0$ have 
all evolved from regions that were at least as big as the second largest spike
found by S98 at $z\sim3$.
This result is consistent with the scenario that the LBOs of $z\sim 3$ 
now reside in rich clusters of galaxies.

Future observations might affect these conclusions in a variety of
ways. It is important to confirm the existence of empty pixels 
in the galaxy distribution as is predicted by all the models. 
With full redshift information for all the galaxies in a data set,
one will be able to use smaller redshift bins (as S98 did in analyzing 
their highest peak) and probably draw stronger conclusions.
It will be interesting to confirm the predictions of a shallow slope 
for the correlation function and a correlation length similar to that 
of nearby galaxies,
and to verify whether it is related to sheet-like versus filamentary structure.

{\bf \noindent Note added:}
Several papers have appeared on this topic since we submitted this
paper; here we briefly comment on how they relate to our work.

Jing \& Suto (1998) evaluate the spike probability using simulations of
three different cosmological models.
Our simulations have comparable force resolution to theirs,
but a higher mass resolution by an order of magnitude. 
They find that a spike the size of the largest one identified by S98 is 
about twice as probable in SCDM (about one in ten fields) compared to our
results (6\% probability per field).
Their published version agrees with our result that the spike
probability in the SCDM model is relatively insensitive to normalization.
They also agree with us that the spikes are more probable than SCDM
in the low-Omega open and flat models they consider, but these results
are not directly comparable to ours since the cosmological parameters of
their models are different from ours.

Governato et al. (1998) use N-body simulations, in which they identify
galaxies with the help of a semi-analytic model, to investigate the
clustering of LBOs.  They do not impose the mean density of LBOs, but rather
use their semi-analytic model to determine whether a given halo has an object
that could be seen in a survey of the type done by S98.  The number densities
that they find from this method are close to that found by S98.
In qualitative agreement with our result, they
find that spikes in the distribution of LBOs like the one observed by
S98 arise naturally in the two models they consider: SCDM and an open CDM
with slightly different parameters than ours.
They show that these spikes become rich clusters in the local
universe, in agreement with our finding.

Peacock (1998) uses a semi-analytic approximation to determine the
bias of LBOs by generating a synthetic redshift histogram and then
estimating the variance in cells the size of S98.  He again finds that
as long as the LBOs are sufficiently biased with respect to the
underlying dark matter, most current CDM-type models can account for
the data.

Moscardini et al. (1998) discuss theoretical predictions for the
number density and correlation function of LBOs.  Their semi-analytic
predictions for the correlation function are qualitatively similar to
ours, and indicate that a better observational determination of the
correlation function may be able to discriminate between models.

It is encouraging that there is a general agreement between the results
of these different investigations using very different methods,
including different $N$-body codes, several different semi-analytic
methods, different ways of identifying halos and galaxies, etc.
The different papers complement each other and provide a coherent
picture with significant confidence.

Giavalisco et al. (1998) calculate the angular correlation function of
871 LBO candidates in five fields of data, and calculate the real-space
correlation function
using the Limber transform.  The correlation lengths that they find are lower
than those we find for any of the models we have considered (see Table 2).
In future work, we plan to identify LBOs in simulations using semi-analytic
models; this may affect the correlation length and power-law index
for all of our models.

\acknowledgments{We have greatly benefited from conversations with
Sandy Faber, Rachel Somerville, James Bullock and Chuck Steidel.  
This research was supported in part by a NASA theory grant and an NSF
theoretical physics grant at UCSC, by the US-Israel Bi-national Science
Foundation grant 95-00330, and by the Israel Science Foundation grant 950/95.
}

\appendix
\section{APPENDIX: SPIKE PROBABILITY AND THE POWER SPECTRUM}

{\footnotesize
\begin{deluxetable}{ccccccccc}
\tablenum{3}
\tablecolumns{9}
\tablecaption{}
\tablehead{
\colhead{Model}                 &
\colhead{$\sigma_8$} &
\colhead{$\sigmap$}            &
\colhead{$\Mh$}         &
\colhead{$\sigmah$}    &
\colhead{$\nu$} &
\colhead{$b$}  &
\colhead{$\sigmahp$}  &
\colhead{$\sigmahp$ ({\rm sim})}\\
\colhead{}&
\colhead{}&
\colhead{}&
\colhead{($\hmsun$)}&
\colhead{}&
\colhead{}&
\colhead{}&
\colhead{}&
\colhead{}}
\startdata
SCDM&   0.67&   0.19& $1.4\times10^{12}$ & 0.73& 2.3& 3.6&   0.68& 0.74\nl
CCDM&   1.27&   0.36& $6.0\times10^{12}$ & 1.06& 1.6& 1.1&   0.69& 0.70\nl
CHDM&   0.72&   0.20& $4.0\times10^{11}$ & 0.62& 2.7& 4.8&   0.98& 0.94\nl
\enddata
\tablecomments{The values of $\sigma$ have been calculated for these
$\Omega=1$ models by using the local
power spectrum and then extrapolating back to high redshift by multiplying
by the scale factor $1/(1+z)$.  For CHDM this is a further approximation,
since the power spectrum shape does change slightly over this range of $z$.}
\end{deluxetable}
}

The dependence of the spike probability on the cosmological parameters
$\Omega$ and $\Lambda$ is qualitatively understood in terms of
the fluctuation growth rates and the comoving volumes of the pixels.
It is not so obvious, however, why the spike probability is found to be
higher in the CHDM model compared to the CDM model, even though they are
both of the same Einstein-deSitter cosmology. It is also not obvious
a priori why the spike probability is found to be relatively insensitive
to the global amplitude (normalization) of fluctuations within a given model,
say CDM.
We offer here a simple framework in which to understand these trends
that we find in the simulations. We provide a heuristic explanation
for the dependence of spike probability on the shape and normalization
of the power spectrum, involving the issues of halo biasing and abundance
(cf. Adelberger \etal 1998).

For each model, we characterize the linear power spectrum of fluctuations
by $\sigma(M)$, the rms linear density fluctuation in top-hat spheres
encompassing a mean mass $M$.
Let $\Mh$ be the galactic-halo mass threshold chosen to reproduce
the correct number density of LBOs; it depends on the
cosmological model, but in all cases it corresponds to linear scales
of order $\sim 1$ $\hmpc$.
Denote $\sigmah\equiv \sigma(\Mh)$.
On the larger scale of the pixels, let $\sigmahp$ and $\sigmap$ be
the rms fluctuations for halos and the underlying matter respectively.
The pixel scale corresponds to a sphere of a comoving radius
$R \simeq 7.6$ $\hmpc$ for $\Omega = 1$.

One way to define a biasing parameter on the pixel scale is via
$b \equiv \sigmahp/\sigmap$.
When the halos are identified as rare peaks in a Gaussian field,
the biasing parameter is approximated by (Mo \& White 1996)
\begin{equation}
b\simeq 1+ \frac{\nu^2-1}{\delc},
\label{eq:mo}
\end{equation}
where $\nu\equiv\delc/\sigmah$ measures the ``rareness'' of the
peaks, and $\delc\simeq 1.686$ is the linear density
threshold for collapsed spherical halos.
In the case of extremely
rare peaks $b \rightarrow \nu/\sigmah$ (Kaiser 1984).
Thus,
\begin{equation}
\sigmahp
\simeq \frac{\sigmap}{\sigmah}\,\nu + \sigmap(1-\delc^{-1}).
\label{eq:sighp}
\end{equation}
The second term is small, $\simeq 0.08$ for SCDM and CHDM, so the
interesting dependencies are in the first (Kaiser) term.
The ratio $\sigmap/\sigmah$ should clearly depend on
the shape of the power spectrum in the sense that a power spectrum with
less power on small compared to large scales
would tend to lead to a higher $\sigmahp$.
This ratio may also depend on normalization indirectly,
because $\Mh$, and thus $\sigmah$,
vary from model to model in order to provide the desired
number density of halos.
The value of $\nu$ varies from model to model for the same reason,
and it could, in principle, depend on shape and normalization.

We can predict for each model the quantities $\Mh$ and $\sigmah$
(or $\nu$), and thus $\sigmahp$.
One relation between $\Mh$ and $\sigmah$ is provided by the power
spectrum itself, $\sigma(M)$.
An independent constraint on these two quantities is imposed by
the fixed number density of halos, enforced to match the observed number
density of LBOs. The halo number density is predicted by the Press--Schechter
approximation to be
\begin{equation}
n_{\rm h}(M) dM  = \sqrt{\frac{2}{\pi}}
\left ( -\frac{{\rm d}\ln\sigma}{dM} \right )
\frac{\bar{\rho}}{M^2} \, \nu\,
e^{(-\nu^2/2)}\, d M.
\label{eq:PS}
\end{equation}

For a given power spectrum $\sigma(M)$, we can integrate equation \ref{eq:PS}
over all masses above the lower threshold $\Mh$ and solve for
$\Mh$ and $\sigmah=\delc/\nu$. Using equation~\ref{eq:mo}, we can then
compute $b$ and $\sigmahp$.

The analytic results for $\sigmahp$ are presented in Table 3
in comparison with the empirical results from the simulations for three
models: (1) SCDM, of $\sigma_8=0.67$ today; (2) CCDM, a COBE-normalized
CDM power spectrum
of about twice the amplitude, $\sigma_8=1.27$; and (3)
our CHDM model, which has a more steeply decreasing power spectrum in the
relevant range of wave-numbers. Shown also
are the predictions for the matter fluctuations on the pixel scale,
$\sigma_p$, the analytic solutions for $\Mh$ and $\sigmah$,
and the corresponding $\nu$ and $b$.

One can see that the predicted halo fluctuations $\sigmahp$,
which directly relate to the high-spike probabilities,
deviate from the simulation values by only 8\% or less.
Although the values predicted for $M_h$ in the analytic model
are systematically  higher than those found in the simulations,
the effect on the final results are small, and the qualitative trends are clear.
The predicted values for the biasing parameter are slightly higher than,
but within the errors of, the values given in \S 5 from the simulations.

What have we learned from the analytic results?
The predicted values for $\sigmahp$ in the CDM model are indeed quite
insensitive to the normalization.
With the higher normalization, there are more halos above
any given mass, so the fixed number density $n_{\rm h}$ requires a larger
threshold $\Mh$.  This naturally reduces the increase in $\sigmah$
due to the higher normalization compared to the increase in $\sigmap$
($\propto \sigma_8$), and therefore leads to a larger $\sigmap/\sigmah$.
However, this is compensated (in equation~\ref{eq:sighp})
by the fact that $\nu$ gets smaller when $M_h$ is larger
(equation~\ref{eq:PS}).

The spike probability for the steeper, CHDM, spectrum is indeed higher
than for SCDM.
Here, there are fewer halos above any given mass, so $M_h$ must go down
in order to keep $n_{\rm h}$ fixed. This naturally corresponds  to an increase
in $\nu$ (equation~\ref{eq:PS}).
Despite the decrease in $M_h$, the shape effect forces $\sigma_h$
to be smaller in CHDM and therefore $\sigmap/\sigmah$ is larger.
The two effects thus both contribute (in equation~\ref{eq:sighp})
to the increase in $\sigmahp$.

Indeed, SCDM has long been known to have too much small-scale power to
match the observed universe; we now see why such a power
spectrum also fails to match the observed clustering of LBOs.
A model with a more realistic power spectrum, such as CHDM or $\tau$CDM,
will also have a higher, more realistic spike probability.
\pagebreak

\end{document}